\documentclass[12pt,a4paper]{panl}
\usepackage{cite}
\usepackage{wrapfig}
\usepackage{graphicx}
\usepackage{amssymb}
\usepackage{amsfonts}
\usepackage{amsmath}
\usepackage{longtable}
\usepackage{rotating}
\usepackage{lscape}
\usepackage{epsfig}
\usepackage{multirow}

\usepackage[english]{babel}
\originalTeX
\begin{document}
\issuearea{Physics of Elementary Particles and Atomic Nuclei. Theory}

\title{Multiplicity correlations in the model with string clusters in pp collisions at LHC energies  
 }
\maketitle
\authors{S.\,Belokurova$^{a,}$\footnote{E-mail: sveta.1596@mail.ru},
V.\,Vechernin$^{a,}$\footnote{E-mail: v.vechernin@spbu.ru}}
\from{$^{a}$\,Saint Petersburg State University, 7/9 Universitetskaya nab., St.	Petersburg, 199034 Russia.}

\begin{abstract}

In the framework of the model with string fusion  and the formation of string clusters the correlations between multiplicities in two separated rapidity windows in pp collisions at LHC energies were studied and
the results obtained were compared with data from the ALICE collaboration at CERN.

The simulation is carried out within the framework of a Monte Carlo implementation of the colour quark-gluon string model. String fusion effects are taken into account by implementing a finite lattice in the plane of the impact parameter.

The dependence of the correlation coefficient between multiplicities in two rapidity observation windows on the distance between these windows is calculated for four values of their width and three values of initial energy.

It is shown that the model with string clusters describes the main features of the behavior of the correlation coefficient: its increase with increasing initial energy, decrease with increasing rapidity distance between observation windows, and nonlinear dependence on the width of the rapidity window.

\end{abstract}
\vspace*{6pt}


\label{sec:intro}
\section*{Introduction}


It is known that the study of correlations and fluctuations of various observables in multiparticle
production processes at high energies gives information about the initial stage of hadronic interaction
corresponding to the maximum density of formed quark–gluon matter.

This study was carried out within the framework of quark-gluon (colour) string models \cite{Kaidalov 1982, Capella 1994} --- a phenomenological model that is used to quantitatively describe soft range of the strong interaction with low momentum transfer, where perturbative QCD calculations are impossible. 
The concept of strings represented by colour flux tubes, which is qualitatively substantiated in the QCD framework, is used in this approach \cite{Casher 1974, Cea 2017}.

This model assumes that interaction occurs in two stages. At the first stage tubes filled with a colour gluon field (strings) are formed as a result of colour reconnection processes. At the second stage observable hadrons are formed as a result of hadronization of these strings.

In the case of high string density in the transverse plane, for example, with
nucleus-nucleus collisions and/or ultra-high energies of the LHC, it is necessary
take into account the interaction between the strings \cite{Biro 1984, Bialas 1986}. For the case of interaction of heavy nuclei, M.A. Brown and C. Pajares proposed \cite{Braun 1992, Braun 1993} the model of fusion (percolation) of primary strings before their fragmentation as a way to take into account the processes of interaction between them.

Later, the simple discrete version of this model was proposed to simplify the consideration of string fusion processes. In this simple model the transverse plane is divided into cells with an area on the order of the cross-sectional area of the string (the cross-sectional radius of the string is $ r_{str} \sim 0, 2 \div 0, 3 $ fm). If the centres of two strings are in the same cell, we assume that they fuse.

\label{sec:corr coef}
\section*{Correlation coefficients and strongly intensive variables}

The study of so-called forward-backward (FB) correlations between observed quantities for particles in two rapidity intervals separated by an interval has been proposed as one of the tools for studying string fusion processes.

Usually these two rapidity intervals (observation windows) are chosen symmetrically, one in the front and the other in the back hemisphere of the reaction.
As observable quantities, $ F $ and $ B $, we usually choose the multiplicity (number) of particles, $ n_{F} $ and $ n_{B} $, with rapidity within the selected intervals,
and the average transverse momentum of these particles in a given event, $p_t^F$ and $p_t^B$,
\begin{equation}
	p_t^F = \frac{1}{n_F}\sum_{i=1}^{n_F} \left| \textbf{p}_t^i\right| ,\ 
	p_t^B = \frac{1}{n_B}\sum_{i=1}^{n_B} \left| \textbf{p}_t^i\right|.
\end{equation}
It was proposed \cite{ALICE 2006} to study 3 types of correlations: $n$ -- $n$, correlations between multiplicities of charged particles,
$p_t$ -- $p_t$, correlations between the average transverse momenta and
$p_t$ -- $n$, correlations between the average transverse momentum in one rapidity window and the multiplicity of charged particles in another rapidity window.

Unfortunately, the coefficient including multiplicity depends on so-called volume fluctuations, which arise due to fluctuations in the volume of the interaction region (for example, the number of strings formed) from event to event. One of the possible ways to exclude the contribution of “volume”
fluctuations and obtaining information about the properties of objects formed at the initial stage of strong interaction is the use of strongly intensive variables \cite{Gorenstein 2011} --- such variables, the value of which does not depend either on the volume of interaction or on its fluctuations.

General methods for constructing such observables were studied in \cite{Gorenstein 2011}. In particular, it was shown that within a certain class of statistical models considered in the work, the value
\begin{equation}\label{strongly intensive}
	\Sigma(A,B)\equiv\frac{\left\langle  A\right\rangle \,\omega_B+\left\langle  B\right\rangle \,\omega_A-2\, \mathrm{cov}(A,B)}{\left\langle  A\right\rangle +\left\langle  B\right\rangle } \ ,
\end{equation}
composed of any two extensive quantities $A$ and $B$ is a strongly intensive variable. In this formula $\left\langle A\right\rangle $ and $\left\langle B\right\rangle $ --- average values of $A$ and  $B$,  $\omega_A$ and $\omega_B$  - their scaled variances:
\begin{equation}\label{scaled variances}
	\omega_A\equiv\frac{D_ A}{\left\langle  A\right\rangle }=\frac{\left\langle  {A^2}\right\rangle -\left\langle  {A}\right\rangle ^2}{\left\langle  A\right\rangle }, 
	\omega_B\equiv\frac{D_ B}{\left\langle  B\right\rangle }=\frac{\left\langle  {B^2}\right\rangle -\left\langle  {B}\right\rangle ^2}{\left\langle  B\right\rangle } \ ,
\end{equation}
$\mathrm{cov}(A,B)$ --- covariance of this variables:
\begin{equation}
	\mathrm{cov}(A,B)=\left\langle  {AB}\right\rangle -\left\langle  {A}\right\rangle \left\langle  {B} \right\rangle  .
\end{equation}

Later it was proposed \cite{Andronov 2015} to use the values of the number of particles $ n_{F} $ and $ n_{B} $, registered in this event in two separated rapidity intervals of these particles in the observation windows usually called forward  and backward, as these extensive variables $ A $ and $ B $.

We will restrict ourselves to the case of symmetric reactions and the equal in width, $ \delta y_F=\delta y_B=\delta y $,
symmetrically located relative to $ y = 0 $ rapidity observation intervals. In this case
\begin{equation} 
	\left\langle n_F \right\rangle =\left\langle n_B\right\rangle =\left\langle n\right\rangle ,\ 
	D_{n_F} = D_{n_B} = D_{n}  ,\ 
	\omega_{n_F} = \omega_{n_B} = \omega_{n} = \frac{D_n}{\left\langle n\right\rangle }.
\end{equation}
Then \eqref{strongly intensive} can be rewritten in a simpler form:
\begin{equation} \label{Sigma}
	\Sigma\left[ n_F,\ n_B\right]  = \frac{ \left\langle n^2\right\rangle - \left\langle n_F\ n_B\right\rangle }{  \left\langle n\right\rangle}.
\end{equation}

For the case of a symmetric reaction and symmetrically located windows of equal width, the following relationship can be obtained between the correlation coefficient $b_{nn}$ and the strongly intensive variable $\Sigma\left[ n_F,n_B\right] $:
\begin{equation}\label{b nn}
	b_{nn}(\Delta y)=1-\frac{\Sigma[n_F,n_B](\Delta y)}{\omega_n}.
\end{equation}
The formula \eqref{b nn} was used to analyse the dependence of the correlation coefficient $b_{nn}$ on the rapidity distance between observation windows.

\section*{Calculation of variables $ \Sigma\left[ n_F,\ n_B\right] $ and $ b_{nn} $ on the lattice}

Within the framework of a model with a finite lattice on a transverse plane, it was shown \cite{Belokurova 2019} that $ \Sigma $ is equal to the sum of the variables $ \Sigma_{\eta} $ introduced for a string cluster in one cell with some weighting coefficients $ \alpha( \eta) $.
\begin{equation}\label{Sigma lat}
	\Sigma[n_F,n_B]
	= \sum_{\eta=1}^{\infty} \alpha(\eta)  \Sigma_\eta (\mu_F, \mu_B)  ,\ 
	\Sigma_\eta (\mu_F, \mu_B)=1+\left\langle {\mu}\right\rangle _{\eta} [J^{\eta}_{FF}-J^{\eta}_{FB}]  .
\end{equation}
The weight coefficients are the average fraction of particles formed as a result of cluster decay.
$ J_{FF}^{\eta} $ and $ J_{FB}^{\eta} $ are expressed through the standard introduced single-particle distribution function of particles (hadrons) $ \lambda_\eta(y) $ formed during the fragmentation of such a string and the two-particle (pair) correlation function $ \Lambda_\eta(y_1,y_2) $, characterizing the correlations between particles formed from the decay of a given string \cite{Belokurova 2019}:
	\begin{equation}\label{JFB}
		J^{\eta}_{FB}\equiv\frac{1}{\left\langle  {\mu_F}\right\rangle _{\eta}\left\langle  {\mu_B}\right\rangle _{\eta}}  \int_{\delta y_F}dy_1 \int_{\delta y_B} dy_2\
		\lambda_\eta(y_1)\lambda_\eta(y_2)\Lambda_\eta(y_1,y_2)    ,
	\end{equation}
	\begin{equation}\label{JFF}
		J^{\eta}_{FF}\equiv\frac{1}{\left\langle  {\mu_F}\right\rangle ^2_{\eta}}  \int_{\delta y_F}dy_1 \int_{\delta y_F} dy_2\
		\lambda_\eta(y_1)\lambda_\eta(y_2)\Lambda_\eta(y_1,y_2)   .
\end{equation}
Similarly, we can obtain the following expression for the scaled variance of the  multiplicity $ \omega_n $ in terms of $ \omega_\mu^\eta $ introduced for the string cluster
\begin{equation}\label{omega n}
	\omega_n = \sum_{\eta=1}^{\infty} \alpha(\eta) \, \omega_\mu^\eta+ \sum_{C_\eta} P(C_\eta) \frac{ \left\langle n\right\rangle _{C_{\eta}}^2  }{\left\langle n\right\rangle }-\left\langle n\right\rangle,\ 
	\omega_{n_i}^{\eta_i} =1+\left\langle n_i\right\rangle_{\eta_i}J_{FF}^{\eta_i}.
\end{equation}
Here $ \left\langle n_i\right\rangle_{\eta_i}=\mu_0\delta y\sqrt{\eta_i}$ is the average multiplicity of charged particles produced during the decay of a cluster of $ \eta_i $ strings in $ i $ -th cell in the rapidity interval $ \delta y $ \cite{Braun 2000}.

If the observation windows $\delta y_F$ and $\delta y_B$ are chosen in the central rapidity region so all the resulting strings contribute to both of these rapidity intervals at once,
then, due to the locality of the strong interaction in the space of rapidities and the resulting homogeneity of the distribution of particles from string fragmentation by rapidity, translational invariance with respect to rapidity takes place.
This approximation works well at LHC energies, when the strings contribute over a wide range of rapidities.
If the two-particle (pair) correlation function $\Lambda_\eta (\Delta y )$ was chosen in its simplest form
\begin{equation}
	\label{corr function}
	\Lambda_\eta (\Delta y ) = \Lambda_0^\eta e^{-\frac{|\Delta y|}{y_{corr}^{(\eta)}}},
\end{equation}
then, within the framework of the assumptions described above, the integrals $J_{FF}^{\eta}$ and $J_{FB}^{\eta}$ can be calculated explicitly:
\begin{equation}\label{J FF}
	J_{FF}^{\eta}
	=\frac{2\Lambda_0^{\eta}}{(\delta y)^2} y_{corr}^{{(\eta)}} \left( \delta y-y_{corr}^{(\eta)}\left( 1-e^{-\frac{ \delta y}{y_{corr}^{{(\eta)}}}}\right)  \right),
\end{equation}
\begin{equation} \label{J FB}
	J_{FB}^{\eta}
	=\frac{\Lambda_0^{\eta} (y_{corr}^{{(\eta)}})^2 }{(\delta y)^2} e^{-\frac{ \Delta y}{y_{corr}^{{(\eta)}}} }\left( e^{-\frac{ \delta y}{y_{corr}^{{(\eta)}}}} +e^{\frac{ \delta y}{y_{corr}^{{(\eta)}}}} -2 \right).
\end{equation}

According to \cite{Vechernin 2018} we assumed the following dependence of the string cluster correlation function parameters on $\eta$:
\begin{equation}\label{y Lambda}
	y_{corr}^{(\eta)} = \frac{y^{(1)}}{\sqrt{\eta}},\ 
	\Lambda_0^{(\eta)} = const.
\end{equation}

\section*{Parameters selection}

Previously \cite{Belokurova Vechernin 2023,Belokurova 2022} the value $\mu_0 = 0.7$ was chosen to describe the multiplicity of all particles with impulses $0<p_t<\infty$. The experimental distribution over $p_t$ gives that it will be 58\% particles in the interval $0.3<p_t<1.5$ GeV, which gives $\mu_0(0.3<p_t<1.5) = 0.7\cdot0.58=0.41$, which was not taken into account earlier.

The parameters \eqref{y Lambda} characterizing clusters with different numbers of fused strings were determined by comparison with experimental data from the ALICE collaboration \cite{Erokhin}:
$ y_{corr}^{(1)} = 2.7,\ \Lambda_0^{(1)} = 0.8 $.

The results for the strongly intensive variable $\Sigma\left( n_F,\ n_B\right) $ depend on the product $\mu_0 \Lambda_0^{(1)}$, therefore, simultaneously with the change in $ \mu_0 $ as a result of taking into account the distribution of particles over $ p_t$, it is also necessary to change the value of the parameter $ \Lambda_0^{(1)}$ so that the product remains the same.
We adjust the parameters as follows:
\begin{equation}\label{new param}
	\mu_0(0.3<p_t<1.5) = 0.41, \
	y_{corr}^{(1)} = 2.7,\
	\Lambda_0^{(1)} = 1.36.
\end{equation}
The \eqref{new param} parameters were used for further calculations.

\section*{Results}

To calculate the value $\Sigma\left[ n_F,n_B\right] $ included in \eqref{b nn}, its expression in terms of $\Sigma_\eta$ \eqref{Sigma lat} was used.
To calculate the scaled variance $\omega_n$ on the lattice, the formula \eqref{omega n} was used.
Thus, the calculation requires only the values of $\eta_i$ --- the number of strings in each cell of the lattice.

To calculate the weighting coefficients $ \alpha(k) $, we first simulated the distribution of the primary strings in the transverse planes taking into account real pp-collision conditions using the Monte Carlo algorithm developed in \cite{Belokurova Vechernin 2023,Belokurova 2022} according to the methodology proposed in the work \cite{Vechernin Lakomov 2012}. Next, the process of fusion of primary strings and forming string clusters was simulated introducing a finite lattice in the impact parameter plane.

We used the following parameter values at the  modelling the distribution of strings in the plane of the impact parameter:
\begin{equation}
	\label{ourparam}
	\Delta= 0.2 \, , \alpha'=0.05\ GeV^{-2} , \,
	\gamma_{pp}=1.035\ GeV^{-2},\,    R_{pp}^2=3.3\ GeV^{-2}\ ,\, C=1.5 .
\end{equation}

\begin{figure}[t]
	\centering
	\begin{tabular}{ccc}
		\hspace{-1cm}		\includegraphics[width=0.38\linewidth]{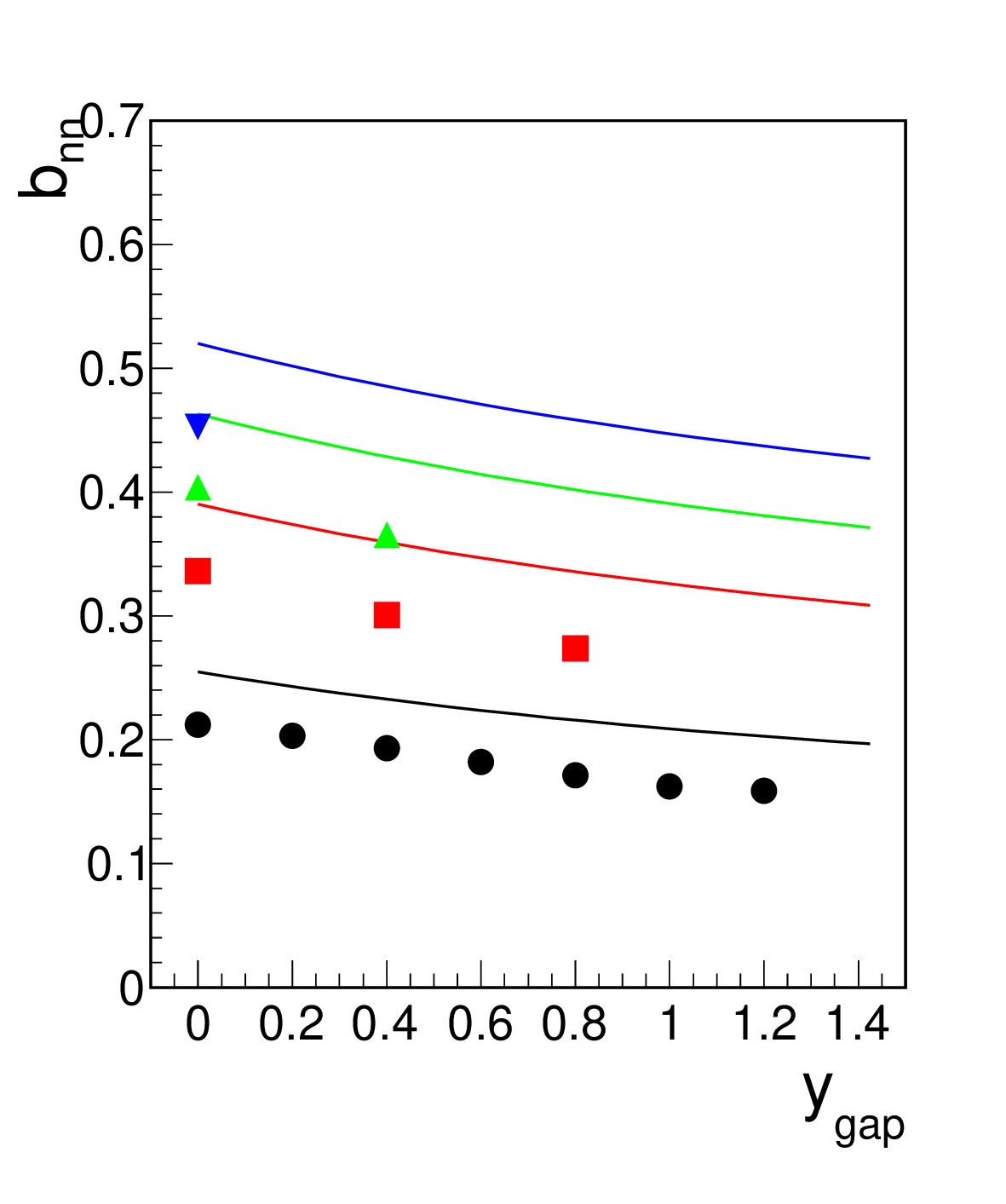}
		\hspace{-1cm}	&\includegraphics[width=0.38\linewidth]{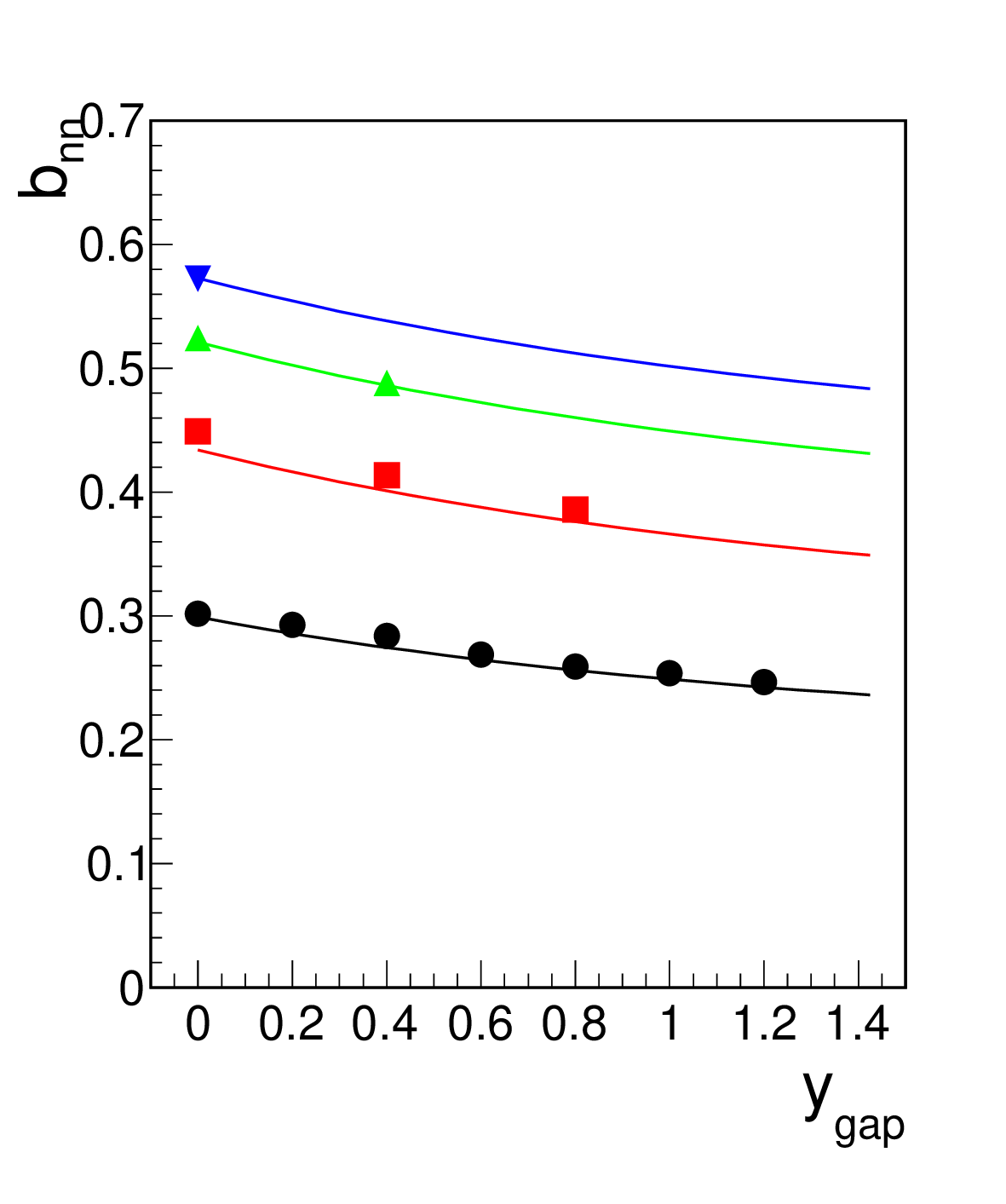} 
		\hspace{-1cm}	&\includegraphics[width=0.38\linewidth]{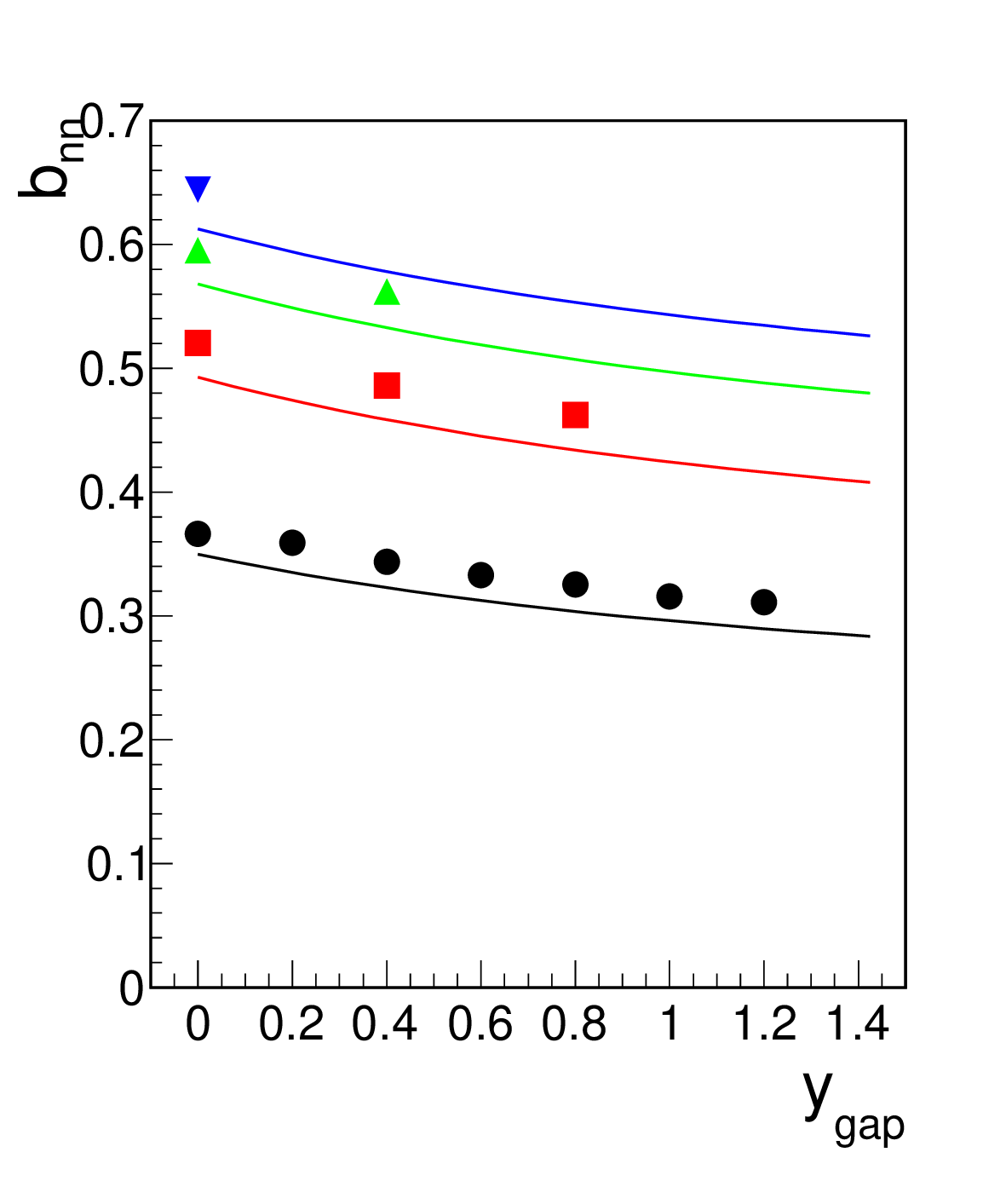} \\ 
	\end{tabular} 
	\caption{Correlation coefficient $b_{nn}$ as a function of the rapidity distance between windows $y_{gap}$ for energies 0.9 TeV (left), 2.76 TeV (center), 7 TeV (right). Lines --- the result of MC calculations for windows with width $\delta y = 0.2, 0.4, 0.6, 0.8$ (from bottom to top). Points --- experimental data obtained by the ALICE collaboration \cite{Erokhin} by analysing data on the yields of charged particles with transverse momenta in the range 0.3-1.5 GeV/$c$ in pp collisions for window widths of 0.2 ($ \bullet$) , 0.4 ($\blacksquare$), 0.6 ($\blacktriangle$), 0.8 ($\blacktriangledown$).
	}
	\label{fig:b MC}
\end{figure}

Fig.\ref{fig:b MC} shows the dependence of the correlation coefficient $b_{nn}$ on the rapidity distance between windows $y_{gap}$ for energies 0.9, 2.76, 7 TeV.

The studied model with the formation of string clusters qualitatively describes the behaviour of $b_{nn}$: a decrease in $b_{nn}$ with increasing distance between windows $y_{gap}$, an increase in $b_{nn}$ with increasing energy, a nonlinear dependence of $b_ {nn}$ from the window width $ \delta y$.

However, we see that the model under study does not provide an ideal numerical description of the absolute value of the correlation coefficient $b_{nn}$. A possible reason is that with the selected parameter values \eqref{ourparam}
our model gives a large value of the scaled variance of the total number of initial strings in the event, which leads to a large value of the scaled variance of the number of charged particles.
Taking into account the processes of string merging significantly reduces the value of the scaled variance of the number of charged particles compared to the model of identical non-interacting strings. Numerically, this reduction allows us to obtain accurate experimental values of $b_{nn}$ only for an initial energy of 2.76 TeV. At an energy of 0.9 TeV this decrease turns out to be insufficient, and at an energy of 7 TeV, on the contrary, it is excessive. To solve this problem, it seems that a revision of the Regge parameters \eqref{ourparam} is necessary. We leave a systematic model analysis of this issue for future research.

\section*{Conclusion}

The previously developed \cite{Belokurova Vechernin 2023,Belokurova 2022} Monte Carlo algorithm was applied to calculate the correlation coefficient between multiplicities $ b_{nn} $ \eqref{b nn} of charged particles within the framework of a model with string fusion on a lattice.

It is shown that the model with string clusters describes the main features of the behaviour of the correlation coefficient $ b_{nn} $: its increase with increasing initial energy, decrease with increasing rapidity distance between observation windows, and nonlinear dependence on the width of the rapidity window. However, it is not possible to obtain an accurate quantitative description of the experimental data for all initial energies.

To solve this problem, we need to revise the Regge parameters \eqref{ourparam}, which are used in our model to find the configurations of the initial strings before taking into account the processes of their fusion. In this case, it is necessary to provide a simultaneous description of a large set of experimental data at different energies --- total, elastic and diffraction cross sections, multiplicities, correlations and fluctuations between various observables. This represents a laborious and important task, which we consider as a subject for our future research.

\section*{Acknowledgments}

Supported by Saint Petersburg State University, project ID: 94031112.

\end{document}